\documentclass[11pt]{article}

\usepackage{fullpage}
\usepackage{graphicx} 
\usepackage{tikz} 
\usepackage{comment} 
\usepackage[tbtags]{amsmath} 
\usepackage{amssymb} 
\usepackage{amsthm} 
\usepackage{mathtools} 
\usepackage{float} 
\usepackage{listings} 
\usepackage{bm} 
\usepackage{subcaption} 
\usepackage{enumitem} 
\usepackage{authblk} 

\usepackage{hyperref} 
\hypersetup{
    colorlinks
}

\usepackage{mleftright} 
\mleftright

\usepackage{physics2} 
\usephysicsmodule{ab} 

\DeclareMathOperator*{\extr}{extr}

\DeclareMathOperator{\tr}{tr}

\DeclareMathOperator{\sgn}{sgn}

\DeclareMathOperator{\erf}{erf}

\DeclareMathOperator{\ReLU}{ReLU}

\DeclarePairedDelimiter{\norm}{\lVert}{\rVert}
\DeclarePairedDelimiter{\ev}{\langle}{\rangle}

\DeclarePairedDelimiterX{\inner}[2]{\langle}{\rangle}{#1, #2}

\newcommand{\de}{\mathrm{d}}

\newcommand{\diffp}[2]{\frac{\partial{#1}}{\partial{#2}}}

\let\R\reals

\let\Z\integers

\newcommand{\idmat}{\mathbf{I}}
\newcommand{\transpose}{\mathsf{T}}
\newcommand{\eps}{\varepsilon}
\newcommand{\relmiddle}[1]{\mathrel{}\middle#1\mathrel{}}

\newcommand{\E}{\mathbb{E}}
\newcommand{\var}{\mathrm{Var}}
\newcommand{\cov}{\mathrm{Cov}}
\newcommand{\normal}{\mathsf{N}}
\newcommand{\Unif}{\mathsf{Unif}}

\newcommand{\pmat}[1]{\begin{pmatrix}#1\end{pmatrix}}

\newcommand{\bw}{\bm{w}}
\newcommand{\bx}{\bm{x}}
\newcommand{\bA}{\bm{A}}
\newcommand{\bB}{\bm{B}}
\newcommand{\bQ}{\bm{Q}}
\newcommand{\bone}{\bm{1}}

\title{Solution Space and Storage Capacity of Fully Connected Two-Layer Neural Networks with Generic Activation Functions}

\author[1]{Sota Nishiyama\thanks{\texttt{snishiyama@g.ecc.u-tokyo.ac.jp}}}
\author[2,3,4]{Masayuki Ohzeki}

\affil[1]{School of Engineering, Tohoku University}
\affil[2]{Graduate School of Information Sciences, Tohoku University}
\affil[3]{Department of Physics, Tokyo Institute of Technology}
\affil[4]{Sigma-i Co., Ltd.}

\begin{document}
\maketitle

\begin{abstract}
    The storage capacity of a binary classification model is the maximum number of random input-output pairs per parameter that the model can learn.
    It is one of the indicators of the expressive power of machine learning models and is important for comparing the performance of various models.
    In this study, we analyze the structure of the solution space and the storage capacity of fully connected two-layer neural networks with general activation functions using the replica method from statistical physics.
    Our results demonstrate that the storage capacity per parameter remains finite even with infinite width and that the weights of the network exhibit negative correlations, leading to a \emph{division of labor}.
    In addition, we find that increasing the dataset size triggers a phase transition at a certain transition point where the permutation symmetry of weights is broken, resulting in the solution space splitting into disjoint regions.
    We identify the dependence of this transition point and the storage capacity on the choice of activation function.
    These findings contribute to understanding the influence of activation functions and the number of parameters on the structure of the solution space, potentially offering insights for selecting appropriate architectures based on specific objectives.
\end{abstract}

\section{Introduction}

\subsection{Background and motivation}

Deep Neural Networks (DNNs) demonstrate remarkable abilities across a wide range of tasks \cite{lecun2015deep}, yet a theoretical explanation for their high performance remains incomplete \cite{berner2021modern,zdeborova2020understanding}.
The theory of deep learning is centered around three problem areas, namely \emph{optimization}, \emph{approximation}, and \emph{generalization} \cite{berner2021modern}.
In the problem of optimization, it is crucial to understand why DNNs can be successfully trained by gradient descent, despite the non-convex nature of their loss functions \cite{auer1995exponentially,safran2018spurious}.
Analyzing the structure of the loss function and the solution space can provide insights into characterizing the condition for successful learning \cite{choromanska2015loss,freeman2016topology,baldassi2019properties}.

The problem of approximation considers the ability of NNs to approximate a given function.
A classical result in this problem is the universal approximation theorem for two-layer NNs \cite{cybenko1989approximation}.
This theorem claims that, under mild assumptions about activation functions, sufficiently wide two-layer neural networks can approximate any continuous function with arbitrary precision.
From a practical standpoint, interest lies in determining precisely how many parameters are needed and in its dependence on the choice of network architectures.

Lastly, the issue of the generalization is about reconciling the bias-variance trade-off and the seemingly excessive number of parameters in modern NNs.
Conventional statistical and learning theories predict a trade-off between the number of parameters in a machine learning model and its generalization performance \cite{hastie2009elements}.
Nonetheless, modern DNNs often contain a large number of parameters, to the extent that they can perfectly fit the training data \cite{zhang2021understanding}.
In particular, the phenomenon of \emph{double descent}, where increasing the number of parameters beyond the capacity to perfectly memorize the dataset further reduces generalization error, has been reported \cite{belkin2019reconciling,geiger2020scaling,nakkiran2021deep}.

Statistical mechanics is a powerful theoretical framework for addressing these problems as a tool to study the typical behavior of systems with many degrees of freedom \cite{zdeborova2020understanding}.
In particular, methods derived from spin glass theory such as replica and cavity methods \cite{mezard1987spin} are useful for investigating the behavior of systems involving randomness and have been applied to evaluate the typical performance of various machine learning models, including NNs \cite{nishimori2001statistical,zdeborova2016statistical,bahri2020statistical}.

In the statistical mechanical analysis of NNs, there are mainly two settings considered.
One is the problem of \emph{storage capacity} \cite{gardner1988optimal,gardner1988space,engel1992storage,barkai1992broken,monasson1995weight,xiong1997storage,baldassi2019properties,zavatone2021activation}.
The storage capacity of a binary classifier refers to the maximum number of random input-output pairs that the model can learn.
A higher storage capacity implies greater representational power and approximation capability of the model.
The analysis of storage capacity examines the typical performance of learning when viewed as an optimization problem or a constraint satisfaction problem.
Among the three problem areas of deep learning theory mentioned earlier, those directly relevant are optimization and approximation.
In addition, one could argue a relationship with the problem of generalization through the phenomenon of double descent.

The other setting is known as the \emph{teacher-student scenario}, which examines the typical performance of learning viewed as a statistical estimation problem \cite{seung1992statistical,watkin1993statistical,biehl1998phase,aubin2018committee,oostwal2021hidden}.
In this setup, random input data and labels generated by a teacher model are provided to a student model for learning, aiming to estimate the parameters of the teacher model.
Subsequently, the generalization ability of the student model to unknown data is evaluated.
This setting is relevant to the problems of optimization and generalization.

These settings together provide complimentary insights for learning of NNs.
In this study, we address the problem of storage capacity.

\subsection{Related works}

The analysis of the storage capacity of NNs using the replica method was pioneered by Gardner and Derrida in their study of perceptrons \cite{gardner1988space,gardner1988optimal}.
They used the replica method to determine the volume of solution space of perceptrons and showed that the storage capacity per parameter of perceptrons is two.

\begin{figure}
    \centering
    \includegraphics[width=0.45\textwidth]{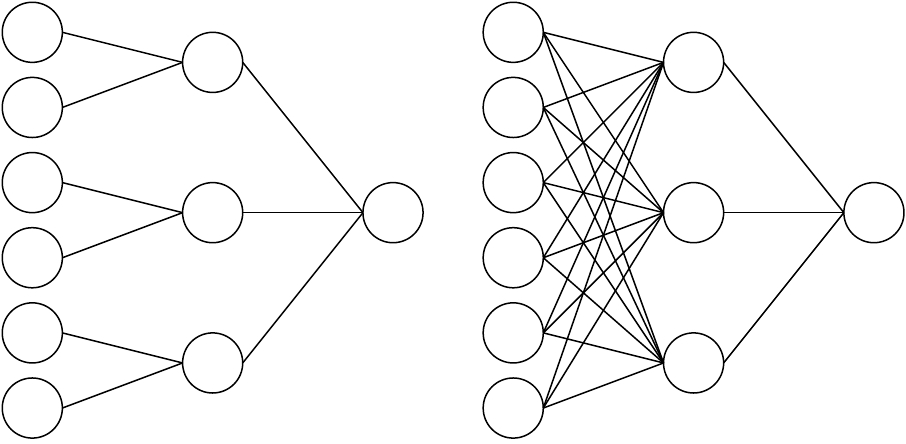}
    \caption{Architectures of a tree-like committee machine (TCM, left) and a fully connected committee machine (FCM, right).}
    \label{fig:committee_arch}
\end{figure}

Gardner's framework was soon extended to two-layer NNs.
Architectures considered in the study of two-layer NNs include tree-like committee machines (TCMs) and fully connected committee machines (FCMs) (see Fig. \ref{fig:committee_arch}).
Storage capacities of TCMs and FCMs with sign activation functions (denoted sgn-TCM and sgn-FCM) were computed in the 1990s \cite{barkai1992broken,engel1992storage,monasson1995weight,xiong1997storage}.
It was demonstrated that in the limit as the width of the hidden layer $K$ tends to infinity, the storage capacity per parameter diverges to infinity as $O(\sqrt{\log K})$.
Also, for sgn-FCMs, it was found that weights connected to the same input neuron have negative correlations, indicating a \emph{division of labor} where different weights attempt to memorize different input-output pairs.

Recently, the storage capacity of TCMs with activation functions other than the sign function, such as Rectified Linear Units (ReLU), has been computed \cite{baldassi2019properties,zavatone2021activation}.
The results showed that, unlike the case of the sign function, the storage capacity remains finite even in the limit as $K$ tends to infinity.

These results are based on the non-rigorous replica method, prompting efforts to support them through more mathematically rigorous approaches.
Stojnic recently published a series of preprints asserting proof that the storage capacities of TCMs derived from replica methods provide rigorous upper bounds on the actual storage capacities \cite{stojnic2023capacity,stojnic2023emph,stojnic2024exact,stojnic2024fixed}.

NNs commonly employed in practical applications often incorporate fully connected layers featuring various activation functions.
Thus, investigating the storage capacity of FCMs with general activation functions, which is yet to be analyzed, would shed light on the performance of NNs used in real-world scenarios.
In this study, we analyze the solution space and the storage capacity of FCMs with general activations, aiming to bridge the gap in previous studies.

\subsection{Organization of this paper}

The remainder of this paper is organized as follows.
In Section \ref{sec:replica}, we formulate the problem of the storage capacity of an FCM and present our analysis using the replica method.
In Section \ref{sec:experiment}, we conduct a numerical experiment where we train FCMs on randomly generated datasets to support our analysis.
Finally, in Section \ref{sec:discussion}, we summarize the results and discuss future directions.

\section{Replica Analysis}
\label{sec:replica}

\subsection{Setup}

The FCM architecture that we consider in this study is a fully connected two-layer network with $N$ inputs, $K$ hidden neurons, and one output.
We make the following assumptions on the activation $g: \R\to\R$: the weak differentiability and the square-integrability of the weak derivative with respect to the Gaussian measure.
Most activation functions used in practice satisfy these conditions, while the sign function does not.

Using the hidden weights $\{\bw_k \in \R^N\}_{k=1,2,\dots,K}$ and the output weights $\{v_k \in \R\}_{k=1,2,\dots,K}$, the output $\hat{y}(\bx) \in \{+1,-1\}$ of the model for an input $\bx \in \R^N$ is given by
\begin{gather}
    \hat{y}(\bx) = \sgn(s(\bx; \{\bw_k\}, \{v_k\})) \,, \label{eq:fcm_def} \\
    s(\bx; \{\bw_k\},\{v_k\}) = \frac{1}{\sqrt{K}} \sum_{k=1}^K v_k g(\bw_k \cdot \bx) \,,
\end{gather}
where $\sgn$ is the sign function.

We fix the output weights $\{v_k\}$ and learn only the hidden weights $\{\bw_k\}$.
To make the distribution of $s$ symmetric even with an asymmetric activation function $g$, we set half of $\{v_k\}$ to $+1$ and the other half to $-1$.
We enforce the norm constraint $\norm{\bw_k}_2^2=N$ for $k=1,2,\dots,K$ on the hidden weights.

The dataset $\mathcal{D}$ is a list of $P$ random input-output pairs $\mathcal{D}=\ab\{(\bx^\mu, y^\mu)\}_{\mu=1,2,\dots,P}$.
Inputs $\bx^\mu \in \R^N$ and outputs $y^\mu \in \{-1,+1\}$ are i.i.d. samples from $\normal(0,N^{-1}\idmat_N)$ and $\Unif(\{-1,+1\})$, respectively, where $\idmat_N$ is the $N\times N$ identity matrix.
The number of samples per parameter is denoted as $\alpha \coloneqq P/(NK)$.

To analyze the solution space structure of FCM, we consider the following partition function or the \emph{Gardner volume}:
\begin{equation}
    Z=\int \prod_{k=1}^K \de\bw_k \, \delta(N-\bw_k\cdot \bw_k) \, \prod_{\mu=1}^P \Theta(y^\mu s(\bx^\mu; \{\bw_k\},\{v_k\})-\kappa) \,,
    \label{eq:partition}
\end{equation}
where $\Theta$ is the Heaviside step function.
This is a volume of the space of weights $\{\bw_k\}$ that classify the dataset $\mathcal{D}$ correctly with a margin of at least $\kappa$.

We expect $\log Z$ to be self-averaging; in the $N \to \infty$ limit, the distribution of $\log Z$ with respect to the realization of the disorder $\mathcal{D}$ concentrates sharply around its mean $[\log Z]_\mathcal{D}$, where $[\cdot]_\mathcal{D}$ is the expected value with respect to the distribution of $\mathcal{D}$.
Under this assumption, we calculate the \emph{quenched free entropy} per parameter
\begin{equation}
    f \coloneqq \lim_{K\to\infty} \lim_{N\to\infty} \frac{1}{NK} [\log Z]_\mathcal{D} \,.
    \label{eq:free_entropy}
\end{equation}
Here, the infinite width limit $K \to \infty$ is taken to simplify the analysis.

\subsection{Replica calculation}

To calculate the quenched free entropy \eqref{eq:free_entropy}, we employ the replica method, which is based on the identity $[\log Z]_\mathcal{D} = \lim_{n\to 0} n^{-1}\log [Z^n]_\mathcal{D}$.
We evaluate $[Z^n]_\mathcal{D}$ for positive integer $n$ and extrapolate the results to $n = 0$.
We do not mathematically justify the analytic continuation from $n\in\Z_+$ to $n\in\R$ and the exchange of limits $n \to 0$ and $N\to\infty$.
The replica method is thus a heuristic rather than a rigorous method.

Here, we sketch the outline of the calculation of $[Z^n]_\mathcal{D}$ and state the main results.
The calculation proceeds along the same line as sgn-FCMs \cite{engel1992storage,barkai1992broken} and TCMs with generic activations \cite{zavatone2021activation}.
We defer the details to the Appendix \ref{app:app}.

\subsubsection{Saddle-point equations}

By evaluating $[Z^n]_\mathcal{D}$ for $n\in \Z_+$, we can formulate $f$ as an extremization problem of a certain objective function as follows:
\begin{equation}
    f=\lim_{n\to 0}\extr_{\bQ,\hat{\bQ} \in \mathcal{S}_{nK}} \ab\{ G_s(\bQ,\hat{\bQ})+\alpha G_e(\bQ) \} \,, \label{eq:free_entropy_generic}
\end{equation}
where we use $\mathcal{S}_{nK}$ to denote the set of $nK \times nK$ positive semi-definite matrices and $\extr_{x \in S} f(x)$ to denote the saddle-point of $f$ i.e. the value $f(x)$ for $x \in S$ such that $f'(x)=0$.
Here, the entropy term $G_s$ and the energy term $G_e$ are given as
\begin{gather}
    G_s(\bQ,\hat{\bQ}) \coloneqq \frac{1}{2nK} \ab(\tr(\bQ\hat{\bQ})-\log\det\hat{\bQ})\,, \\
    G_e(\bQ) \coloneqq \frac{1}{n} \log \ab[ \prod_{a=1}^n \Theta\ab(y\frac{1}{\sqrt{K}}\sum_{k=1}^K v_k g(v_kh_k^a) - \kappa) ]_{\mathcal{D}}\,,
\end{gather}
where
\begin{equation}
    h_k^a \coloneqq v_k \bw_k^a\cdot \bx\,.
\end{equation}
For replica indices $1 \leq a,b \leq n$ and hidden unit indices $1 \leq k,l \leq K$, components of the order parameters $\bQ,\hat{\bQ}$ are defined as
\begin{gather}
    Q_{kl}^{ab} \coloneqq \frac{1}{N} v_kv_l\bw_k^a \cdot \bw_l^b=\begin{cases}
        1      & (k=l, a=b)           \\
        q^{ab} & (k=l, a\neq b)       \\
        c^a    & (k\neq l, a=b)       \\
        d^{ab} & (k \neq l, a\neq b)
    \end{cases} \,,
    \label{eq:rs_order_params} \\
    \hat{Q}_{kl}^{ab} \coloneqq\begin{cases}
        \hat{E}^a     & (k=l, a=b)           \\
        -\hat{q}^{ab} & (k=l, a\neq b)       \\
        \hat{c}^a     & (k\neq l, a=b)       \\
        -\hat{d}^{ab} & (k \neq l, a\neq b)
    \end{cases} \,,
    \label{eq:rs_conj_params}
\end{gather}
where we omit the dependence on the hidden unit indices $k,l$ due to the symmetry of the problem.
$q^{ab}$ represents the overlap of the same weights across different replicas, characterizing the size of \emph{clusters} or the number of solutions.
Here, a cluster refers to a single connected region in the solution space of the FCM.
As $q$ approaches one, the size of clusters decreases and the number of solutions diminishes.

The order parameters $c^a$ and $d^{ab}$ are unique to FCMs and were introduced previously in the study of sgn-FCMs \cite{engel1992storage,barkai1992broken}.
$c^a$ denotes the overlap of different weights within a single replica, characterizing the degree of a division of labor.
A negative value of $c$ suggests the occurrence of a division of labor.
$d^{ab}$ is associated with the concept of \emph{permutation symmetry} (PS).
For an FCM, permuting the indices of the hidden weights of a solution yields another solution.
When this permuted solution belongs to the same cluster as the original solution, we say that PS holds.
PS phase is characterized by $q^{ab}=d^{ab}$.
If solutions belong to different clusters, we say that there is a permutation symmetry breaking (PSB), characterized by $q^{ab} \neq d^{ab}$.
In sgn-FCM, it is known that PS holds for small $\alpha$, but PSB occurs for large $\alpha$ \cite{barkai1992broken}.

Solving the extremization problem of Eq. \eqref{eq:free_entropy_generic} for general $nK\times nK$ positive semi-definite matrices $\bQ,\hat{\bQ}$ is intractable.
Hence we need to make assumptions on the structure of $\bQ,\hat{\bQ}$ and parametrize them with a small number of order parameters.
We consider two kinds of assumptions: \emph{replica symmetry} (RS) and \emph{one-step replica symmetry breaking} (1-RSB).

Under the RS ansatz, we assume that the components of $\bQ$ and $\hat{\bQ}$ do not depend on the replica indices $a,b$ and parametrize them with seven scalar order parameters $q,c,d,\hat{E},\hat{q},\hat{c},\hat{d} \in \R$.
This ansatz is equivalent to assuming that typical solutions of the FCM form a single connected cluster.
Using these order parameters, the free entropy \eqref{eq:free_entropy_generic} is written as
\begin{multline}
    f_{\text{RS}}=\extr_{q,\bar{c},\bar{d},\hat{E},\hat{q},\hat{c},\hat{d}} \Bigg\lbrace \frac{1}{2} \ab( \hat{E} + q\hat{q} + \bar{c}\hat{c} + \bar{d}\hat{d} + \frac{\hat{q}-\hat{d}}{\hat{E}+\hat{q}-\hat{c}-\hat{d}}-\log(\hat{E}+\hat{q}-\hat{c}-\hat{d}) ) \\
    + \alpha \int Dz \log H\ab(\frac{\kappa+\sqrt{\lambda}z}{\sqrt{\mu}}) \Bigg\rbrace \,,
    \label{eq:free_entropy_rs}
\end{multline}
where $Dx$ is the standard Gaussian measure $\exp(-x^2/2)/\sqrt{2\pi}\,\de x$ and $H(x)$ is the tail distribution of the standard normal distribution $H(x)=\int_x^\infty Dz$.
In addition, we have defined the rescaled order parameters $\bar{c} \coloneqq (K-1)c,\,\bar{d} \coloneqq (K-1)d$ and defined $\lambda,\mu$ as
\begin{align}
    \lambda & = q_\text{eff}+\ev{g'}^2\bar{d} \,,           \label{eq:lambda_rs}       \\
    \mu     & = \sigma^2-q_\text{eff}+\ev{g'}^2 (\bar{c}-\bar{d}) \,, \label{eq:mu_rs}
\end{align}
where $\ev{\cdot}$ is the expected value with respect to the standard Gaussian measure $Dx$, $\sigma^2 = \ev{g^2}-\ev{g}^2$ is the variance of $g$, and $q_\text{eff}$ is the \emph{effecive order parameter} defined as
\begin{equation}
    q_\text{eff} \coloneqq \cov\ab[g(x),g(y) \relmiddle| \pmat{x \\ y} \sim \normal\ab(0, \pmat{1 & q \\ q & 1})] \,. \\
    \label{eq:q_eff}
\end{equation}

By differentiating the RS free entropy \eqref{eq:free_entropy_rs} with each order parameter, we obtain saddle-point equations, a system of equations that gives a solution to the extremization problem \eqref{eq:free_entropy_rs}, as follows:
\begin{align}
    \hat{q}                               & = \alpha \frac{q_\text{eff}'}{\mu} \int Dz \, \ab(\frac{H'(u)}{H(u)})^2 \,,
    \label{eq:sp_rs_qhat}                                                                                                 \\
    \hat{c}                               & = \alpha \frac{\ev{g'}^2}{\mu} \int Dz \, \frac{H'(u)}{H(u)} u \,,
    \label{eq:sp_rs_chat}                                                                                                 \\
    \hat{d}                               & = \alpha \frac{\ev{g'}^2}{\mu} \int Dz \, \ab(\frac{H'(u)}{H(u)})^2 \,,
    \label{eq:sp_rs_dhat}                                                                                                 \\
    \hat{E} + \hat{q} - \hat{c} - \hat{d} & = \frac{1 + \sqrt{4(\hat{q}-\hat{d})+1}}{2} \,,
    \label{eq:sp_rs_Ehat}                                                                                                 \\
    q                                     & = \frac{\hat{q}-\hat{d}}{(\hat{E}+\hat{q}-\hat{c}-\hat{d})^2} \,,
    \label{eq:sp_rs_q}                                                                                                    \\
    \bar{c}                               & = -1 \,,
    \label{eq:sp_rs_cbar}                                                                                                 \\
    \bar{d}                               & = -q \,,
    \label{eq:sp_rs_dbar}
\end{align}
where $u=(\kappa+\sqrt{\lambda}z)/\sqrt{\mu}$.

We obtain the result $c=-1/(K-1),d=-q/(K-1)$ regardless of $\alpha$, $\kappa$ and $g$.
Note that $\bar{c}=-1$ is the minimum possible value of $\bar{c}$, since the positive semi-definiteness of the overlap matrix requires $1 + \bar{c} \geq 0$.
Hence, the degree of division of labor is maximized at $\bar{c}=-1$.
In addition, the positive semi-definiteness of the overlap matrix requires $1-q+\bar{c}-\bar{d}\geq 0$ and $q+\bar{d} \geq 0$. Given that $\bar{c}=-1$, then $\bar{d}=-q$ is the only possible solution.

\begin{figure}
    \centering
    \includegraphics{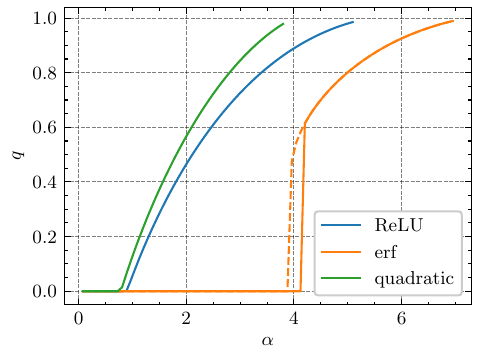}
    \caption{(Color online) The overlap between replicas $q$ at $\kappa=0$ for activations ReLU, erf, and quadratic $x^2$.
        We solved the saddle-point equations \eqref{eq:sp_rs_qhat}--\eqref{eq:sp_rs_dbar} iteratively starting from an appropriate set of initial values.
        For ReLU and quadratic, the transition from the PS phase with $q=0$ to the PSB phase with $q > 0$ is continuous.
        The transition point $\alpha_\text{PS}$ is approximately $0.897$ for ReLU and $0.785$ for quadratic.
        On the other hand, in the case of erf activation, we observe a discontinuous transition to the PSB phase.
        At $\alpha_\text{spin}\approx 3.949$, a local minimum of $f_\text{RS}$ with $q > 0$ appears (dashed line).
        Above $\alpha_\text{PS}\approx 4.142$, the $q>0$ solution gives the global minimum of the $f_\text{RS}$, marking a discontinuous transition to the PSB phase (solid line).
    }
    \label{fig:fcm_overlap}
\end{figure}

The order parameter $q$ can be determined by solving the saddle-point equations \eqref{eq:sp_rs_qhat}--\eqref{eq:sp_rs_dbar}.
We show in Fig. \ref{fig:fcm_overlap} the solutions for three activation functions: the rectified linear unit $\ReLU(x)=\max\{x,0\}$, the error function $\erf(x)=2/\sqrt{\pi} \int_0^x \exp(-t^2) \, \de t$, and the quadratic function $x^2$.
The overlap $q$ increases monotonically as the number of samples per parameter $\alpha$ increases, showing a decrease in the number of solutions.
For small $\alpha$, we obtain a solution with $q=0$, and for large $\alpha$, we obtain a solution with $q>0$.
The transition from the $q=0$ solution to the $q>0$ solution is where the permutation symmetry breaks and will be discussed in the next subsection.
At some $\alpha$, $q$ reaches one and there exists a unique set of weights that correctly classifies the given dataset.
This value of $\alpha$ is the RS storage capacity.

\subsubsection{Permutation symmetry and its breaking}

At the transition point $\alpha_\text{PS}$ which depends on the activation function, the system undergoes a phase transition from the PS phase with $q=d=0$ to the PSB phase with $q \neq d$.
To examine the properties of this transition, we derive the condition for the PS solution $q=d=0$ to be locally stable in the saddle-point equations \eqref{eq:sp_rs_qhat}--\eqref{eq:sp_rs_dbar}.
When $q \ll 1$ and $\hat{q}-\hat{d} \ll 1$, the leading terms in the saddle-point equations are given by
\begin{align}
    \hat{q}-\hat{d} & \approx \alpha \frac{\ev{g''}^2}{\mu} \ab(\frac{H'(u)}{H(u)})^2 q \,, \\
    q               & \approx \hat{q}-\hat{d} \,,
\end{align}
where $u=\kappa/\sqrt{\sigma^2 - \ev{g'}^2}$.
Hence, the condition for $(q,\hat{q}-\hat{d})=(0,0)$ to be locally stable in the RS saddle-point equations can be expressed as follows:
\begin{equation}
    \alpha < \frac{\sigma^2 - \ev{g'}^2}{\ev{g''}^2} \ab(\frac{H(u)}{H'(u)})^2 \,.
    \label{eq:ps_stability}
\end{equation}

For activation functions satisfying $\ev{g''} \neq 0$, the right-hand side of Eq. \eqref{eq:ps_stability} takes a finite value, which we denote as $\alpha_\text{PS0}$.
Such activation functions include ReLU and quadratic functions.
When $\kappa=0$, we have $\alpha_\text{PS0} = (\pi/2-1)\pi/2 \approx 0.897$ for ReLU and $\alpha_\text{PS0} = \pi/4 \approx 0.785$ for quadratic functions.
These values coincide with the point where $q$ rises from zero in Fig. \ref{fig:fcm_overlap}, suggesting that the transition to the PS phase occurs continuously for these activation functions and that the transition point $\alpha_\text{PS}$ satisfies $\alpha_\text{PS}=\alpha_\text{PS0}$.

On the other hand, for activation functions satisfying $\ev{g''}=0$, the condition in Eq. \eqref{eq:ps_stability} is always satisfied, showing that the PS solution is always locally stable.
This suggests that the transition to the PSB phase is a discontinuous first-order transition.
This class of activation functions includes erf and, more generally, any odd function.

The transition to the PSB phase for $g=\erf$ occurs as follows.
As we increase $\alpha$, at the spinodal point $\alpha_\text{spin}\approx 3.949$, a local solution $q\approx 0.41$ appears, distinct from the $q=0$ solution.
At $\alpha=\alpha_\text{spin}$, the value of the free entropy $f_\text{RS}$ for the $q > 0$ solution is greater than that of the $q = 0$ solution.
Then, at $\alpha_\text{PS}\approx 4.142$, $q\approx 0.59$ becomes the global minimum of $f_\text{PS}$, leading to a discontinuous transition from the PS solution with $q=0$ to the PSB solution with $q>0$.

Note that $\ev{g''} \neq 0$ does not necessarily imply a continuous transition to the PS phase. For example, $g(x)=\erf(x) + 0.05x^2$ has a finite $\alpha_\text{PS0} \approx 7.091$ but the transition to the PS phase occurs discontinuously at $\alpha_\text{PS}\approx 3.572$.

\subsubsection{Storage capacity}

The value of $\alpha$ at which $q=1$ corresponds to the storage capacity $\alpha_\text{RS}$.
It can be obtained from the saddle-point equations and is expressed by the following equation:
\begin{equation}
    \frac{1}{\alpha_\text{RS}} = \frac{\sigma^2 - \ev{g'}^2}{\ev{g'^2} - \ev{g'}^2} \int_{-\kappa/\sqrt{\sigma^2 - \ev{g'}^2}}^\infty Dz \, \ab(\frac{\kappa}{\sqrt{\sigma^2 - \ev{g'}^2}}+ z)^2 \,.
    \label{eq:rs_cap}
\end{equation}
In particular, at $\kappa = 0$, we have
\begin{equation}
    \alpha_\text{RS}(\kappa=0) = \frac{2 \ab(\ev{g'^2} - \ev{g'}^2)}{\sigma^2 - \ev{g'}^2} \,.
    \label{eq:rs_cap_zero}
\end{equation}

Omitting $\ev{g'}^2$ terms in these expressions gives the storage capacity of the corresponding TCM \cite{zavatone2021activation}.
$\ev{g'}^2$ can be considered as a term characterizing the effect of division of labor.
A comparison between the RS storage capacities of FCM and TCM is shown in Table \ref{table:alpha_rs_tcm_fcm}.
For activation functions with $\ev{g'}\neq 0$, such as ReLU and erf, the RS storage capacities of FCMs are higher than those of the corresponding TCMs by the division of labor.
However, for activation functions with $\ev{g'}=0$, such as the quadratic function, the division of labor does not increase the storage capacities of FCMs.

\begin{figure}
    \centering
    \includegraphics{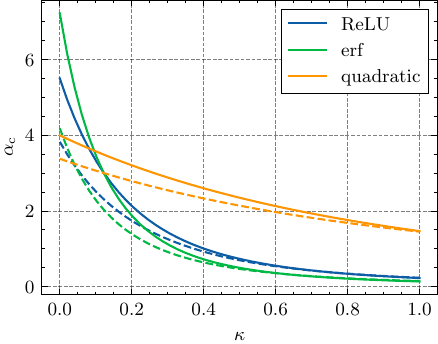}
    \caption{(Color online) Storage capacities of FCMs with ReLU, erf, and quadratic activations.
        Solid lines represent the RS storage capacities and dashed lines represent the 1-RSB storage capacities.
        For small $\kappa$, there is a large gap between RS and 1-RSB storage capacities, showing a strong replica symmetry-breaking effect.
    }
    \label{fig:alpha_1rsb_fcm}
\end{figure}

{
\renewcommand{\arraystretch}{1.8}
\begin{table}
    \caption{Comparison of RS storage capacities of FCMs and TCMs at $\kappa=0$.
        RS storage capacities of FCMs are given by Eq. \eqref{eq:rs_cap_zero}.
        RS storage capacities of TCMs are obtained by setting $\ev{g'}^2$ to zero in Eq. \eqref{eq:rs_cap_zero} \cite{zavatone2021activation}.
        For ReLU and erf, FCM capacity is larger than TCM capacity due to the effect of division of labor.
        For the quadratic function, however, the FCM capacity matches the TCM capacity because $\ev{g'}=0$.
    }
    \label{table:alpha_rs_tcm_fcm}
    \centering
    \begin{tabular}{c|cc}
        \hline
        Activation & FCM capacity                                            & TCM capacity                                   \\
        \hline \hline
        ReLU       & $\dfrac{2\pi}{\pi-2} \approx 5.504$                     & $\dfrac{2\pi}{\pi-1}\approx 2.934$             \\
        erf        & $\dfrac{4(3/\sqrt{5}-1)}{3\arcsin(2/3)-2}\approx 7.223$ & $\dfrac{4}{\sqrt{5}\arcsin(2/3)}\approx 2.451$ \\
        quadratic  & $4$                                                     & $4$                                            \\
        \hline
    \end{tabular}
\end{table}
}

When $\ev{g'}>0$, the RS storage capacity of FCM at $\kappa=0$ exceeds that of TCM.
This holds for cases including ReLU and erf.
When $\ev{g'}=0$, the RS storage capacities of FCM and TCM coincide regardless of $\kappa$.
This holds for cases including quadratic functions.
In general, for an even activation function $g$, we have $\ev{g'}=0$.

\subsubsection{1-RSB calculation and the stability of RS solution}

One can evaluate the free entropy \eqref{eq:free_entropy_generic} under 1-RSB ansatz to obtain the 1-RSB storage capacity.
1-RSB capacities are shown in Fig. \ref{fig:alpha_1rsb_fcm} along with RS capacities.
The explicit values of the 1-RSB capacities $\alpha_\text{1-RSB}$ for several activation functions at $\kappa=0$ are:
$3.812$ for ReLU, $4.186$ for erf, and $3.375$ for the quadratic.

When solving the 1-RSB saddle-point equations, the RS solution is restored for small $\alpha$.
Above a certain threshold $\alpha_\text{AT}$, the RS solution becomes locally unstable and the 1-RSB solution emerges continuously.
This instability of the RS solution is known as the de Almeida-Thouless (dAT) instability \cite{de1978stability}.

To determine $\alpha_\text{AT}$, we examine the condition under which the 1-RSB saddle-point equations have $(\Delta q,\Delta {\hat{q}},\Delta {\bar{d}},\Delta {\hat{d}})=(0,0,0,0)$ as a stable fixed point, where $\Delta q\coloneqq q_1-q_0,\,\Delta {\hat{q}}\coloneqq \hat{q}_1-\hat{q}_0,\,\Delta {\bar{d}}\coloneqq \bar{d}_1-\bar{d}_0,\,\Delta {\hat{d}}\coloneqq \hat{d}_1-\hat{d}_0$.
Thus, the condition for dAT stability for a given $(\alpha,\kappa)$ can be obtained as follows:
\begin{equation}
    \alpha (1-q)^2 \int Dz \, \Bigg(\frac{ (q_\text{eff}' - \ev{g'}^2)^2 }{\lambda^2} \ab(\frac{\partial^2}{\partial z^2} \log H(u))^2 + \frac{q_\text{eff}''}{\lambda}\ab(\frac{\partial}{\partial z} \log H(u))^2\Bigg) < 1 \,,
    \label{eq:dat_instability}
\end{equation}
where the values of $q$ and $u$ are solutions to the RS saddle-point equations for the given $(\alpha,\kappa)$.

When we keep only the leading contributions by setting $q\ll 1$ in Eq. \eqref{eq:dat_instability}, the condition coincides with Eq. \eqref{eq:ps_stability}.
Hence, for activation functions where PSB solutions appear continuously, $\alpha_\text{PS}$ and $\alpha_\text{AT}$ coincide.
In other words, the PSB phase and the RSB phase coincide.

Moreover, when evaluating the stability condition \eqref{eq:dat_instability} for the RS solution with $\kappa=0$ and $g=\erf$, we find that PS solutions for $\alpha < \alpha_\text{PS}\approx 4.142$ are stable, while PSB solutions for $\alpha > \alpha_\text{spin}\approx 3.949$ are unstable.
Thus, erf also exhibits simultaneous transition to RSB and PSB.

\subsubsection{Comparison with teacher-student scenario}

The PSB transition discussed in this section is analogous to a phenomenon known as \emph{specialization transition} in the analysis of generalization errors of FCM in the teacher-student scenario.
The specialization transition occurs when the weights of the student's FCM specialize to mimic specific weights of the teacher's FCM, breaking the symmetry of weight permutations.
The specialization transition is known to be discontinuous for sgn and erf, while it is continuous for ReLU \cite{seung1992statistical,biehl1998phase,aubin2018committee,oostwal2021hidden}.
It is noteworthy that the continuity of the PSB transition and that of the specialization transition match, at least for ReLU and erf.

\section{Numerical Experiment}
\label{sec:experiment}

The Gardner volume solely addresses the existence of solutions, without guaranteeing the presence of algorithms capable of efficiently discovering solutions in regions where they exist.
Notably, training neural networks are known to be NP-hard in general \cite{blum1988training,vsima2002training}, suggesting the absence of polynomial-time algorithms for solution discovery.
Consequently, empirically verifying the accuracy of the storage capacity derived through the replica method via numerical experiments presents a formidable challenge.
Nonetheless, the discovery of solutions surpassing the theoretically derived storage capacity by an algorithm would serve to refute the theory.
Thus, through experimental evidence demonstrating the inability of two-layer neural networks trained via gradient descent to discover solutions beyond the storage capacity for random datasets, we indirectly corroborate the validity of our calculation.

\subsection{Settings of experiments}

We implement the FCM of Eq. \eqref{eq:fcm_def} using PyTorch \cite{paszke2019pytorch} and train it using gradient descent on randomly generated datasets.
We fix the number of parameters $NK$ to $10000$ and vary the ratio $N/K$ over $\{0.5, 1, 5, 10, 50\}$.
Since the replica analysis assumes $N \gg K$, higher values of $N/K$ are closer to the settings of replica calculations.
We examine the cases of ReLU and erf as activation functions $g$.
The parameter $\alpha$ is varied from $0.2$ to $4.8$ in increments of $0.2$.

We use hinge loss with $\kappa=0$ and the Adam optimizer \cite{kingma2014adam} for the learning algorithm.
We terminate the training when the loss reaches below $10^{-5}$ or when the training reaches 5000 epochs.

For each combination of $(\alpha, N/K, g)$, we train the model on 10 different datasets.
We compute the accuracy, loss, number of epochs until training completion, and the value of weight overlap $c$ at the end of training for each trial and then average them over the 10 trials.

Although we do not measure the order parameters $q$ and $d$ representing the overlaps across different replicas due to a limited amount of computational resources,
they correspond to measurable physical quantities and can be computed in principle: by preparing multiple FCMs corresponding to different replicas, training them on the same input data, and calculating the overlap of the weights between different FCMs.

\subsection{Results}

\begin{figure}
    \centering
    \includegraphics[width=\textwidth]{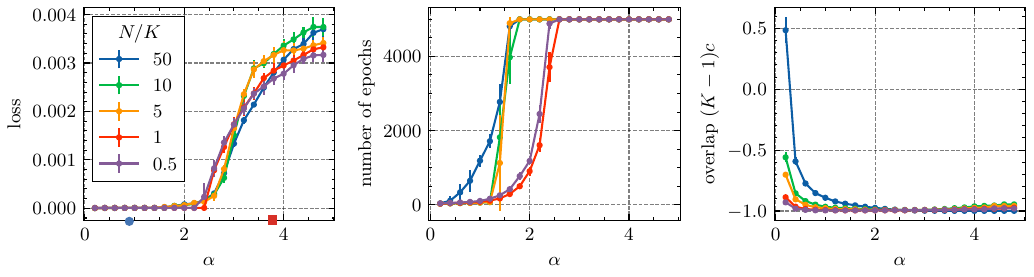}
    \caption{(Color online) Results for ReLU.
        We plot the value of the loss at the end of training (left), the number of epochs to the end of training (middle), and the overlap of the hidden weights (right).
        The blue dot and the red square on the axis of the leftmost figure indicate the theoretical values of $\alpha_\text{PS}$ and $\alpha_\text{1-RSB}$, respectively.
        We observe a sudden increase in the loss at around $\alpha=2.5$ and estimate it to be the experimental storage capacity of FCM with ReLU activations.
        The number of training epochs until termination increases rapidly as $\alpha$ grows.
        The overlap between weights $c$ reaches $-1/(K-1)$ for sufficiently large $\alpha$, consistent with the replica analysis.
    }
    \label{fig:relu_result}
\end{figure}

The results for ReLU are shown in Fig. \ref{fig:relu_result}.
We observe abrupt increases in the loss values at around $\alpha \approx 2.5$, consistently across the values of $N/K$.
Thus, the experimental storage capacity for ReLU is estimated to be around 2.5, which is significantly smaller than the 1-RSB storage capacity.
This discrepancy is likely due to the breaking of replica symmetry and the emergence of numerous local minima at large $\alpha$, causing gradient descent to become trapped in local minima.
Indeed, the replica analysis predicts the breaking of replica symmetry in the PSB phase $\alpha > \alpha_\text{PS} \approx 0.897$, suggesting the highly nonconvex loss landscape with numerous local minima.
A similar phenomenon is observed in the numerical experiment for TCMs \cite{zavatone2021activation}.
Hence, the gap in the experimental storage capacity and the theoretical 1-RSB storage capacity does not falsify the theory, as long as the former is upper-bounded by the latter.

In addition, the number of epochs required for training completion increases sharply as $\alpha$ grows larger.
Also, the overlap $(K-1)c$ converges to $-1$ as $\alpha$ increases, which is consistent with the prediction of the replica analysis.
The increase in $c$ at small $\alpha$ is attributed to finite-size effects.
Convergence to $-1$ is faster for larger values of $K$.

\makeatletter
\newcommand*\fsize{\dimexpr\f@size pt\relax}
\makeatother

\begin{figure}
    \centering
    \includegraphics[width=\textwidth]{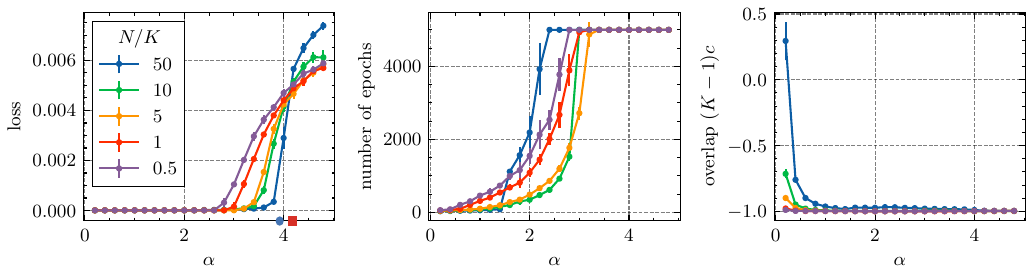}
    \caption{(Color online) Results for erf.
        The loss behaves differently for different input-hidden ratios $N/K$.
        The blue dot and the red square on the axis of the leftmost figure indicate the theoretical values of $\alpha_\text{spin}$ and $\alpha_\text{1-RSB}$, respectively.
        For $N/K=50$, the experimental storage capacity seems to be around 3.8 and this is well explained by our theory as stated in the main text.
        Since the replica analysis assumes infinite $N/K$, it fails to describe the behavior for small $N/K$.
        As with ReLU, the number of epochs until termination increases rapidly as $\alpha$ increases, and consistent with the theory, the order parameter $c$ converges to $-1/(K-1)$.
    }
    \label{fig:erf_result}
\end{figure}

Next, the results for erf are presented in Fig. \ref{fig:erf_result}.
For erf, the loss does not seem to behave consistently across different $N/K$.
For $N/K=50$, which is the closest to the setting of replica analysis, the experimental storage capacity is estimated to be around 3.8, close to $\alpha_\text{spin}\approx 3.9$ at which a PSB solution appears.
Above $\alpha_\text{spin}$, the replica symmetry is broken and the gradient descent algorithm is likely to get trapped in a local minimum.
This explains the gap between the experimental storage capacity and the 1-RSB storage capacity $\alpha_\text{1-RSB}\approx 4.186$.

\section{Discussion}
\label{sec:discussion}

\subsection{Summary of results}

We have analyzed the structure of the solution space of fully connected two-layer neural networks (FCMs) with general activation functions using the replica method and determined their storage capacity.
Under mild assumptions on the activation functions, we have shown that the storage capacity remains finite even in the infinite width limit $K \to \infty$.
This result differs from those of sgn-TCM and sgn-FCM while resembling the results obtained for TCM with general activation functions.
Also, the observation that the order parameter $c$ becomes negative indicates the occurrence of a division of labor, similar to the sgn-FCM.

In addition, the boundary between the PS and PSB phases, denoted as $\alpha_\text{PS}$, is identified.
The transition to the PSB phase is found to be continuous when ReLU or quadratic functions are used as activation functions, while it is discontinuous when erf is used.

Finally, we have conducted a numerical experiment to train FCMs using gradient descent.
In these experiments, the gradient descent algorithm is only able to find solutions up to significantly smaller $\alpha$ values compared to the 1-RSB storage capacity.
This discrepancy is attributed to the non-convexity of the solution space, leading to a gap between the existence of solutions and their learnability.
However, consistent with our theory, the experimental estimate of the storage capacity of erf is found to be greater than that of ReLU.
In addition, the value of $c$ matched the theoretical prediction of $-1/(K-1)$.

The practical significance of this study can be summarized as follows.
First, it enables the comparison of neural network performance based on the choice of activation functions.
By observing the differences in storage capacity ($\alpha_\text{RS}, \alpha_\text{1-RSB}$) and the transition points to PSB/RSB ($\alpha_\text{PS}, \alpha_\text{AT}$) across different activation functions, appropriate choices of activation functions depending on the task and computational resources can be made.
Second, it aids in the design of architectures with appropriate parameter sizes.
Understanding the impact of parameter sizes on the structure of the solution space allows for determining the necessary $\alpha$ to train models in the over-parameterized regime or the RS/PS regions.

\subsection{Perspectives}

Gardner's framework of storage capacity examines the uniform distribution on the solution space, which solutions obtained by learning algorithms such as gradient descent do not generally follow.
In particular, gradient descent is expected to perform some form of implicit regularization; that is, it converges to solutions with some desirable properties among the set of possible solutions, leading to solutions with high generalization performance when applied to over-parameterized neural networks \cite{neyshabur2014search}.
Thus, the analysis of the dependency of the solution space structure and the storage capacity on learning algorithms is an important future task.

In addition, it is necessary to consider datasets with some structure, rather than random data, to understand practical machine learning.
As an attempt to handle datasets with structure within the framework of statistical mechanics, Goldt et al. proposed the \emph{hidden manifold model} \cite{goldt2020modeling}.
It should be possible to calculate the storage capacity of FCMs using datasets following the hidden manifold model.

Exploring the relationship between FCM's storage capacity and the correspondence with the neural network-Gaussian process (NN-GP) is also significant.
Zavatone-Veth and Pehlevan pointed out that the effective order parameter $q_\text{eff}$ characterizing the storage capacities of TCMs formally coincides with the kernel of a Gaussian process corresponding to a two-layer neural network \cite{zavatone2021activation}.
The storage capacity of FCMs, however, is determined from $\lambda = q_\text{eff}-\ev{g'}^2q$ in Eq. \eqref{eq:lambda_rs}, not just $q_\text{eff}$.
One can argue a correspondence between $\lambda$ and the kernel of a two-layer neural network with its weights initialized to have mutually negative correlations.
The practical significance of such an initialization scheme awaits verification in future research.

Lastly, given that many practical neural networks are deep with three or more layers, identifying the structure of the solution space and storage capacity of DNNs is of high importance.
In particular, the dependence of the storage capacity on the number of layers is one of the most interesting aspects.
To this end, several results regarding the storage capacity of DNNs have been reported.
Geiger et al. experimentally determined the storage capacity of DNNs trained on binary classification tasks using random and real datasets and found that the storage capacity does not depend on the number of layers \cite{geiger2019jamming}.
In addition, Yoshino calculated the storage capacity of DNNs with all layers of width fixed at a constant value $N$ using the replica method and found that the storage capacity increases exponentially with the number of layers \cite{yoshino2020complex}.
We note that Yoshino's results should not apply to DNNs with a single output neuron;
for the storage capacity per parameter $\alpha_c$ and the Vapnik-Chervonenkis dimension per parameter $\alpha_\text{VC}$ of a binary classifier, it is proved that $\alpha_c < 2\alpha_\text{VC}$ holds \cite{engel2001statistical} and that $\alpha_\text{VC}$ is of at most polynomial order of the number of layers \cite{bartlett2019nearly}.
Investigating Geiger et al.'s results about the storage capacity of DNNs with one output or a constant number of outputs theoretically is an important future task.
We expect that Yoshino's method as well as a series of studies on the storage capacity of two-layer neural networks including this study will be useful.

\section*{Acknowledgements}
We thank Yoshihiko Nishikawa and Manaka Okuyama for helpful discussions and comments.
This work is supported by JSPS KAKENHI Grant No. 23H01432.
Our study receives financial support from the programs for Bridging the gap between R\&D and the IDeal society (society 5.0) and Generating Economic and social value (BRIDGE) and the Cross-ministerial Strategic Innovation Promotion Program (SIP) from Cabinet Office.

\bibliographystyle{jpsj}
\bibliography{reference}

\newpage

\appendix
\setcounter{equation}{0}
\renewcommand{\theequation}{\thesection.\arabic{equation}}

\section{Details of replica calculation}
\label{app:app}

\subsection{Derivation of the quenched free entropy}

In this subsection, we calculate the replicated partition function
\begin{equation}
    [Z^n] = \int \prod_{k=1}^K\prod_{a=1}^n \de \bw_k^a \, \delta(N-\bw_k^a \cdot \bw_k^a)
    \,\ab[\prod_{a=1}^n \Theta(ys(\bx; \{\bw_k^a\},\{v_k\}) - \kappa)]_{\bx,y}^P \,.
    \label{eq:replicated_partition_function}
\end{equation}
Define the local fields as $h_k^a=v_k \bw_k^a \cdot \bx$.
Since the components of $\bx$ are i.i.d. normal random variables, their weighted sums $\{h_k^a\}_{k=1,2,\dots,K}^{a=1,2,\dots,n}$ follow a multivariate normal distribution.
Its mean and covariance are given by
\begin{align}
    \E[h_k^a]         & =0 \,,                                              \\
    \cov[h_k^a,h_l^b] & =\frac{1}{N}v_kv_l \bw_k^a \cdot \bw_l^b \,.
\end{align}
Introducing the order parameters as defined in the main text, the replicated partition function is written as follows:
\begin{align}
    \begin{split}
        [Z^n] & \propto \int \prod_{ka} \ab( \de {\bw_k^a} \, \delta(N-\bw_k^a \cdot \bw_k^a) ) \prod_{a<b} \ab( \de{q^{ab}} \, \prod_k \delta(Nq^{ab}-\bw_k^a \cdot \bw_k^b) )                           \\
        & \quad\times \prod_{a} \ab( \de{c^{a}} \, \prod_{k<l} \delta(Nc^{a}-v_kv_l\bw_k^a \cdot \bw_l^a) ) \prod_{a<b} \ab( \de{d^{ab}} \, \prod_{k<l} \delta(Nd^{ab}-v_kv_l\bw_k^a \cdot \bw_l^b) ) \\
        & \quad\times \ab[ \prod_a \Theta\ab(y\frac{1}{\sqrt{K}} \sum_kv_kg(v_kh_k^a) - \kappa) ]^P_{h,y}
    \end{split} \\
    \begin{split}
        & = \int \prod_{a<b} \de{q^{ab}} \prod_{a} \de{c^a} \prod_{a<b} \de{d^{ab}} \, \ab[ \prod_a \Theta\ab(y\frac{1}{\sqrt{K}} \sum_kv_kg(v_kh_k^a) - \kappa) ]^P_{h,y}  \\
        & \quad \times \Bigg( \int \prod_{ka} \de{\bw_k^a} \, \prod_{ka} \delta(N-\bw_k^a \cdot \bw_k^a) \prod_{a<b,k} \delta(Nq^{ab}-\bw_k^a \cdot \bw_k^b) \\
        & \quad\qquad \times \prod_{a,k<l} \delta(Nc^{a}-v_kv_l\bw_k^a \cdot \bw_l^a) \prod_{a<b,k<l} \delta(Nd^{ab}-v_kv_l\bw_k^a \cdot \bw_l^b) \Bigg) \,.
    \end{split}
    \label{eq:replicated_partition_function_with_ord}
\end{align}

The expectation can be written as
\begin{equation}
    \ab[ \prod_a \Theta\ab(y\frac{1}{\sqrt{K}} \sum_kv_kg(v_kh_k^a) - \kappa) ]^P_{h,y} = \exp(nPG_e(\bQ)) = \exp(\alpha nNK G_e(\bQ)) \,,
\end{equation}
where we defined the energy term as
\begin{equation}
    G_e(\bQ) \coloneqq \frac{1}{n} \log \ab[ \prod_a \Theta\ab(y\frac{1}{\sqrt{K}} \sum_kv_kg(v_kh_k^a) - \kappa) ]_{h,y} \,.
\end{equation}

Next, we evaluate the integral in the second line of Eq. \eqref{eq:replicated_partition_function_with_ord}.
Using the integral representation of the delta function
\begin{equation}
    \delta(x)=\int_{-\infty}^\infty \frac{\de{k}}{2\pi} \exp(ikx)
\end{equation}
and the Gaussian integral, in the $n\to 0$ limit, the integral is calculated as
\begin{equation}
    \int \de{\hat{\bQ}} \, \exp\ab(\frac{N}{2}\ab(\tr(\bQ\hat{\bQ}) - \log\det\hat{\bQ})) = \int \de{\hat{\bQ}} \, \exp\ab(nNK G_s(\bQ,\hat{\bQ})) \,,
\end{equation}
using the conjugate order parameter $\hat{\bQ}$ introduced in the main text.
Here, we defined the entropy term as
\begin{equation}
    G_s(\bQ,\hat{\bQ}) \coloneqq  \frac{1}{2nK} \ab(\tr(\bQ\hat{\bQ}) - \log\det\hat{\bQ}) \,.
\end{equation}

Combining, Eq. \eqref{eq:replicated_partition_function_with_ord} becomes
\begin{equation}
    [Z^n] \propto \int \de{\bQ} \de{\hat{\bQ}} \, \exp\ab(nNK \ab(\alpha G_e(\bQ) + G_s(\bQ,\hat{\bQ}))) \,.
\end{equation}
In the $N\to\infty$ limit, applying the saddle-point method gives
\begin{equation}
    [Z^n] \propto \exp\ab(nNK \extr_{\bQ,\hat{\bQ}} \ab\{\alpha G_e(\bQ) + G_s(\bQ,\hat{\bQ})\}) \,.
\end{equation}
Thus, the quenched free entropy is
\begin{equation}
    f = \extr_{\bQ,\hat{\bQ}} \ab\{G_s(\bQ,\hat{\bQ}) + \alpha G_e(\bQ)\} \,.
\end{equation}

\subsection{The free entropy and the storage capacity under RS ansatz}

From now on, we assume $c,d$ are of $O(1/K)$.
It will be shown that this assumption is self-consistent.
Define the rescaled order parameters $\bar{c},\bar{d}$ as
\begin{equation}
    \bar{c} \coloneqq (K-1)c,\quad \bar{d} \coloneqq (K-1)d \,.
\end{equation}

First, we evaluate the entropy term $G_s(\bQ,\hat{\bQ})$.
Using the formula
\begin{equation}
    \det \underbrace{\pmat{ \bA & \bB & \cdots & \bB \\ \bB & \bA & & \vdots \\ \ \vdots & & \ddots & \bB \\ \bB & \dots & \bB & \bA }}_{n} = \det(\bA+(n-1)\bB)\ab(\det(\bA-\bB))^{n-1} \,,
\end{equation}
we obtain
\begin{align}
    \begin{split}
        \det \hat{\bQ} & = \ab(1 - n\frac{\hat{q}+(K-1)\hat{d}}{\hat{E}+\hat{q}+(K-1)(\hat{c}+\hat{d})} ) \times \ab(\hat{E}+\hat{q}+(K-1)(\hat{c}+\hat{d}))^n    \\
        & \quad \times \ab(1-n\frac{\hat{q}-\hat{d}}{\hat{E}+\hat{q}-\hat{c}-\hat{d}} )^{K-1} \times \ab(\hat{E}+\hat{q}-\hat{c}-\hat{d})^{(K-1)n} \,,
    \end{split} \\
    G_s(\bQ,\hat{\bQ}) & = \frac{1}{2nK} \ab(\tr(\bQ\hat{\bQ}) - \log \det \hat{\bQ})                                                                                                                                                                                                                                                             \\
    \begin{split}
        & \xrightarrow{n\to 0} \frac{1}{2}\ab(\hat{E} + q\hat{q} + (K-1)(c\hat{c} + d\hat{d}))                                                          \\
        & \qquad - \frac{1}{2K}\ab(\log(\hat{E}+\hat{q}+(K-1)(\hat{c}+\hat{d})) - \frac{\hat{q}+(K-1)\hat{d}}{\hat{E}+\hat{q}+(K-1)(\hat{c}+\hat{d})} ) \\
        & \qquad - \frac{K-1}{2K} \ab(\log(\hat{E}+\hat{q}-\hat{c}-\hat{d}) - \frac{\hat{q}-\hat{d}}{\hat{E}+\hat{q}-\hat{c}-\hat{d}})
    \end{split}                                                                                  \\
                   & \xrightarrow{K \to \infty}
    \frac{1}{2}\ab(\hat{E} + q\hat{q} + \bar{c}\hat{c} + \bar{d}\hat{d} - \ab(\log(\hat{E}+\hat{q}-\hat{c}-\hat{d}) - \frac{\hat{q}-\hat{d}}{\hat{E}+\hat{q}-\hat{c}-\hat{d}})) \,.
\end{align}

Next, we evaluate the energy term $G_e(\bQ)$. Let
\begin{equation}
    s^a \coloneqq \frac{1}{\sqrt{K}} \sum_k v_k g(v_k h_k^a) \,.
\end{equation}
Although $s^a$ is a sum of a large number of random variables, $\{h_k\}_{k=1,\dots, K}$ have non-zero pairwise covariance $c$ and thus are not independent.
Hence, the central limit theorem is not directly applicable.
However, since the covariance $c$ is small (of $O(1/K)$), one can still approximate $s^a$ with a normal random variable.
Therefore, $\{s^a\}_{a=1,\dots,n}$ follow a multivariate normal distribution with mean and covariance given by:
\begin{align}
    \E[s^a]       & = 0 \,,                                                                                          \\
    \cov[s^a,s^b] & =\frac{1}{K}\sum_{kl} v_kv_l\cov[g(v_kh_k^a),g(v_lh_l^b)]                                    \\
                  & = \cov[g(h_k^a),g(h_k^b)] + (K/2-1)\cov[g(h_k^a),g(h_l^b)] - (K/2)\cov[g(h_k^a),g(-h_l^b)] \,.
\end{align}
Now, define the effective order parameters as follows:
\begin{align}
    \sigma^2         & \coloneqq \var[g(x) \mid x \sim \normal(0,1)] \,,
    \label{eq:rs_sigma}                                                 \\
    q_\text{eff}     & \coloneqq \cov\ab[g(x),g(y) \relmiddle| \pmat{x \\ y} \sim \normal\ab(0, \pmat{1 & q \\ q & 1})] \,,
    \label{eq:rs_qtilde}                                                \\
    c_\text{eff}^\pm & \coloneqq \cov\ab[g(x),g(y) \relmiddle| \pmat{x \\ y} \sim \normal\ab(0, \pmat{1 & \pm c \\ \pm c & 1})] \,,
    \label{eq:rs_ctilde}                                                \\
    d_\text{eff}^\pm & \coloneqq \cov\ab[g(x),g(y) \relmiddle| \pmat{x \\ y} \sim \normal\ab(0, \pmat{1 & \pm d \\ \pm d & 1})] \,.
    \label{eq:rs_dtilde}
\end{align}
Using the assumption that $c,d$ are of $O(1/K)$ and expanding $g$ to the first order, we obtain
\begin{align}
    c_\text{eff}^\pm & = \pm \ev{g'}^2 c + O(1/K^2) \,,  \\
    d_\text{eff}^\pm & = \pm \ev{g'}^2 d + O(1/K^2) \,.
\end{align}
Therefore,
\begin{equation}
    \cov[s^a,s^b] = \delta_{ab}\ab(\sigma^2+\ev{g'}^2\bar{c}) + (1-\delta_{ab})\ab(q_\text{eff}+\ev{g'}^2\bar{d}) + O(1/K) \,.
\end{equation}
Introducing $\lambda$ and $\mu$ as
\begin{align}
    \lambda & \coloneqq q_\text{eff}+\ev{g'}^2\bar{d} \,,                      \\
    \mu     & \coloneqq \sigma^2-q_\text{eff}+\ev{g'}^2 (\bar{c}-\bar{d}) \,,
\end{align}
$s^a$ can be written as
\begin{equation}
    s^a = \sqrt{\lambda}z + \sqrt{\mu} x^a \,,
\end{equation}
where $x^1,\dots,x^n,z \sim \normal(0,1)$ i.i.d.
Combining these results, the energy term becomes
\begin{align}
    G_e(Q) & = \frac{1}{n} \log \ab[ \prod_a \Theta\ab(ys^a - \kappa) ]_{s,y}                                       \\
           & = \frac{1}{n} \log \ab[ \int Dz \, \ab( \int Dx \, \Theta(y(\sqrt{\lambda}z+\sqrt{\mu}x)-\kappa))^n]_{y} \\
           & = \frac{1}{n} \log \int Dz \, \ab( H\ab(\frac{\kappa+\sqrt{\lambda}z}{\sqrt{\mu}}))^n                 \\
           & \xrightarrow{n\to 0} \int Dz \, \log H\ab(\frac{\kappa+\sqrt{\lambda}z}{\sqrt{\mu}}) \,.
    \label{eq:rs_energy}
\end{align}
Thus we obtain the RS free entropy given in Eq. \eqref{eq:free_entropy_rs}.

To obtain the saddle-point equations, we differentiate the free entropy \eqref{eq:free_entropy_rs} with each order parameter.
First, we derive the equation for $q$. Let
\begin{equation}
    \mathcal{Y} = H\ab(\frac{\kappa + \sqrt{\lambda}z}{\sqrt{\mu}}) = \int Dx \, \Theta\ab(\kappa + \sqrt{\lambda} z + \sqrt{\mu}x) \,.
\end{equation}
Then,
\begin{align}
    \diffp{f_\text{RS}}{q} & = \frac{1}{2} \hat{q} + \alpha \int Dz \, \diffp{\mathcal{Y}}{q}                                                                                                                                                                           \\
                           & = \frac{1}{2} \hat{q} + \alpha \int Dz \, \frac{1}{\int Dx \, \Theta} \int Dx \, \Theta' \cdot \ab(\frac{z}{2\sqrt{\lambda}}-\frac{x}{2\sqrt{\mu}}) q_\text{eff}'                                                                        \\
                           & = \frac{1}{2} \hat{q} + \alpha q_\text{eff}' \int Dz \, \ab( \frac{z}{2\sqrt{\lambda}} \diffp{\mathcal{Y}}{(\sqrt{\lambda}z)} - \frac{1}{2} \ab(\frac{\partial^2\mathcal{Y}}{\partial (\sqrt{\lambda}z)^2} + \ab(\diffp{\mathcal{Y}}{(\sqrt{\lambda}z)})^2 )) \\
                           & = \frac{1}{2} \hat{q} - \frac{1}{2}\alpha q_\text{eff}' \int Dz \, \ab(\diffp{\mathcal{Y}}{(\sqrt{\lambda}z)})^2 \,,
\end{align}
where $\partial/\partial(\sqrt{\lambda}z)$ means $(1/\sqrt{\lambda})\partial/\partial z$.
This gives the saddle-point equation shown in Eq. \eqref{eq:sp_rs_qhat}.
Similar calculation for $\bar{c},\bar{d}$ gives the saddle-point equations shown in Eq. \eqref{eq:sp_rs_chat} and \eqref{eq:sp_rs_dhat}.

The derivative with respect to $\hat{E}$ is
\begin{equation}
    \diffp{f_\text{RS}}{\hat{E}} = \frac{1}{2} - \frac{1}{2} \frac{\hat{E}-\hat{c}+2(\hat{q}-\hat{d})}{(\hat{E}+\hat{q}-\hat{c}-\hat{d})^2} \,.
\end{equation}
Solving this for $\hat{E}+\hat{q}-\hat{c}-\hat{d}$ gives
\begin{equation}
    \hat{E} + \hat{q} - \hat{c} - \hat{d} = \frac{1 + \sqrt{4(\hat{q}-\hat{d})+1}}{2} \,.
\end{equation}
Differentiating with respect to $\hat{q},\hat{c},\hat{d}$ gives the saddle-point equations shown in Eq. \eqref{eq:sp_rs_q}, \eqref{eq:sp_rs_cbar}, and \eqref{eq:sp_rs_dbar}.

Next, we evaluate the RS storage capacity $\alpha_\text{RS}$.
Let $q=1-\eps$ and consider the limit $\eps \to 0$.
By the weak differentiability and the square-integrability of the weak derivative with respect to the Gaussian measure, the effective order parameter can be expanded as $q_\text{eff}=\sigma^2 - \ev{g'^2}\eps + O(\eps^2)$ \cite{zavatone2021activation}.
Thus,
\begin{equation}
    \lambda \to \sigma^2 - \ev{g'}^2 \,,\quad
    \mu = \ab(\ev{g'^2} - \ev{g'}^2)\eps + O(\eps^2) \eqqcolon \gamma \eps + O(\eps^2) \,,
\end{equation}
and $u=(\kappa + \sqrt{\lambda}z)/\sqrt{\mu} = O(\eps^{-1}) \gg 1$.
Using the asymptotic form of $H(x)$
\begin{equation}
    H(x) = \begin{cases}
        \frac{1}{\sqrt{2\pi} x} \exp\ab(-\frac{x^2}{2}) \ab(1 + O(x^{-2}))     & (x \gg 1)   \\
        1 - \frac{1}{\sqrt{2\pi} x} \exp\ab(-\frac{x^2}{2}) \ab(1 + O(x^{-2})) & (x \ll -1)
    \end{cases} \label{eq:H_asym} \\
\end{equation}
and $H'(x)=-\exp(-x^2/2)/\sqrt{2\pi}$, Eq. \eqref{eq:sp_rs_qhat} is
\begin{align}
    \hat{q} & \to \alpha \frac{\ev{g'^2}}{\mu} \int_{-\kappa/\sqrt{\lambda}}^\infty Dz \, \ab(\frac{\kappa + \sqrt{\lambda}z}{\sqrt{\mu}})^2     \\
            & = \alpha \frac{\ev{g'^2}\lambda}{\gamma^2 \eps^2} \int_{-\kappa/\sqrt{\lambda}}^\infty Dz \, \ab(\frac{\kappa}{\sqrt{\lambda}}+z)^2 \,.
\end{align}
Eq. \eqref{eq:sp_rs_dhat} is
\begin{align}
    \hat{d} & \to \alpha \frac{\ev{g'}^2}{\mu} \int_{-\kappa/\sqrt{\lambda}}^\infty Dz \, \ab(\frac{\kappa + \sqrt{\lambda}z}{\sqrt{\mu}})^2     \\
            & = \alpha \frac{\ev{g'}^2\lambda}{\gamma^2 \eps^2} \int_{-\kappa/\sqrt{\lambda}}^\infty Dz \, \ab(\frac{\kappa}{\sqrt{\lambda}}+z)^2 \,.
\end{align}
Therefore, using $1-q = 1/(\hat{E}+\hat{q}-\hat{c}-\hat{d})$, Eq. \eqref{eq:sp_rs_q} is written as follows:
\begin{align}
    q & = \frac{\hat{q}-\hat{d}}{(\hat{E}+\hat{q}-\hat{c}-\hat{d})^2}                                                 \\
      & = \alpha \frac{\lambda}{\gamma} \int_{-\kappa/\sqrt{\lambda}}^\infty Dz \, \ab(\frac{\kappa}{\sqrt{\lambda}}+z)^2 \,.
\end{align}
This gives the RS storage capacity shown in Eq. \eqref{eq:rs_cap}.

\subsection{The free entropy and the storage capacity under 1-RSB ansatz}

Under the 1-RSB ansatz, we parametrize the overlap matrix $\bQ$ and its conjugate $\hat{\bQ}$ with twelve order parameters $m,q_1,q_0,c,d_1,d_0,\hat{E},\hat{q}_1,\hat{q}_0,\hat{c},\hat{d}_1,\hat{d}_0\in\R$ as follows:
\begin{gather}
    \bQ \coloneqq \underbrace{
        \pmat{
            \bA & \bB & \cdots & \bB \\
            \bB & \bA & & \bB\\
            \vdots & & \ddots & \vdots \\
            \bB & \cdots & \bB & \bA
        }}_{K} \,,
    \quad
    \hat{\bQ} \coloneqq \underbrace{
        \pmat{
            \hat{\bA} & \hat{\bB} & \cdots & \hat{\bB} \\
            \hat{\bB} & \hat{\bA} &  & \hat{\bB} \\
            \vdots &  & \ddots & \vdots \\
            \hat{\bB} & \cdots & \hat{\bB} & \hat{\bA}
        }}_{K} \,,
\end{gather}
\begin{gather}
    \bA \coloneqq \underbrace{
        \pmat{
            \bA_1 & \bA_0 & \cdots & \bA_0 \\
            \bA_0 & \bA_1 & & \bA_0\\
            \vdots & & \ddots & \vdots \\
            \bA_0 & \cdots & \bA_0 & \bA_1
        }}_{n/m} \,,
    \quad
    \bB \coloneqq \underbrace{
        \pmat{
            \bB_1 & \bB_0 & \cdots & \bB_0 \\
            \bB_0 & \bB_1 & & \bB_0\\
            \vdots & & \ddots & \vdots \\
            \bB_0 & \cdots & \bB_0 & \bB_1
        }}_{n/m} \,, \\
    \bA_1 = q_1 \bone_m \bone_m^\transpose + (1-q_1) \idmat_m \,,
    \quad
    \bA_0 = q_0 \bone_m \bone_m^\transpose \,, \\
    \bB_1 = d_1 \bone_m \bone_m^\transpose + (c-d_1) \idmat_m \,,
    \quad
    \bB_0 = d_0 \bone_m \bone_m^\transpose \,,
\end{gather}
\begin{gather}
    \hat{\bA} \coloneqq \underbrace{
        \pmat{
            \hat{\bA}_1 & \hat{\bA}_0 & \cdots & \hat{\bA}_0 \\
            \hat{\bA}_0 & \hat{\bA}_1 & & \hat{\bA}_0\\
            \vdots & & \ddots & \vdots \\
            \hat{\bA}_0 & \cdots & \hat{\bA}_0 & \hat{\bA}_1
        }}_{n/m} \,,
    \quad
    \hat{\bB} \coloneqq \underbrace{
        \pmat{
            \hat{\bB}_1 & \hat{\bB}_0 & \cdots & \hat{\bB}_0 \\
            \hat{\bB}_0 & \hat{\bB}_1 & & \hat{\bB}_0\\
            \vdots & & \ddots & \vdots \\
            \hat{\bB}_0 & \cdots & \hat{\bB}_0 & \hat{\bB}_1
        }}_{n/m} \,, \\
    \hat{\bA}_1 = -\hat{q}_1 \bone_m \bone_m^\transpose + (\hat{E}+\hat{q}_1) \idmat_m \,,
    \quad
    \hat{\bA}_0 = -\hat{q}_0 \bone_m \bone_m^\transpose \,,
\end{gather}
where $\bone_m$ is the $m$-dimensional vector with all components set to one.

Again we self-consistently assume that $c,d_1,d_0$ are of $O(1/K)$, and define the rescaled order parameters as
\begin{equation}
    \bar{c} \coloneqq (K-1)c \,,\quad \bar{d}_1 \coloneqq (K-1)d_1 \,,\quad \bar{d}_0 \coloneqq (K-1)d_0 \,.
\end{equation}

First, we evaluate the entropy term $G_s(\bQ,\hat{\bQ})$.
\begin{align}
    \begin{split}
        G_s & \xrightarrow{n\to 0} \frac{1}{2}\ab(\hat{E}+(1-m)q_1\hat{q}_1+mq_0\hat{q}_0+(K-1)(c\hat{c}+(1-m)d_1\hat{d}_1+md_0\hat{d}_0))                                                                   \\
        & \qquad - \frac{1}{2K} \Bigg( \frac{1}{m} \log \ab( \hat{E}+(1-m)\hat{q}_1+m\hat{q}_0 + (K-1)(\hat{c}+(1-m)\hat{d}_1+m\hat{d}_0))                                                                \\
        & \qquad\qquad + \ab(1-\frac{1}{m}) \log \ab( \hat{E}+\hat{q}_1+(K-1)(c+\hat{d}_1) )                                                                                                                   \\
        & \qquad\qquad - \frac{\hat{q}_0+(K-1)\hat{d}_0}{ \hat{E}+(1-m)\hat{q}_1+m\hat{q}_0 + (K-1)(\hat{c}+(1-m)\hat{d}_1+m\hat{d}_0)} \Bigg)                                                                 \\
        & \qquad - \frac{K-1}{2K} \Bigg( \frac{1}{m} \log \ab( \hat{E}+(1-m)\hat{q}_1+m\hat{q}_0-\hat{c}-(1-m)\hat{d}_1-m\hat{d}_0 ) \\
        & \qquad\qquad + \ab(1-\frac{1}{m}) \log \ab( \hat{E}+\hat{q}_1-\hat{c}-\hat{d}_1 ) \\
        & \qquad\qquad - \frac{\hat{q}_0-\hat{d}_0}{ \hat{E}+(1-m)\hat{q}_1+m\hat{q}_0-\hat{c}-(1-m)\hat{d}_1-m\hat{d}_0 }  \Bigg)
    \end{split} \\
    \begin{split}
        & \xrightarrow{K \to \infty} \frac{1}{2}\ab(\hat{E}+(1-m)q_1\hat{q}_1+mq_0\hat{q}_0+\bar{c}\hat{c}+(1-m)\bar{d}_1\hat{d}_1+m\bar{d}_0\hat{d}_0)                                               \\
        & \qquad - \frac{1}{2} \Bigg( \frac{1}{m} \log \ab( \hat{E}+(1-m)\hat{q}_1+m\hat{q}_0-\hat{c}-(1-m)\hat{d}_1-m\hat{d}_0 ) \\
        & \qquad\qquad + \ab(1-\frac{1}{m}) \log \ab( \hat{E}+\hat{q}_1-\hat{c}-\hat{d}_1 ) \\
        & \qquad\qquad - \frac{\hat{q}_0-\hat{d}_0}{ \hat{E}+(1-m)\hat{q}_1+m\hat{q}_0-\hat{c}-(1-m)\hat{d}_1-m\hat{d}_0 }  \Bigg) \,.
    \end{split}
\end{align}

Next, we evaluate the energy term $G_e(\bQ)$.
Calculation along the same line as the RS ansatz gives
\begin{equation}
    G_e \xrightarrow{n \to 0} \frac{1}{m} \int Dz_0 \, \log \int Dz_1 \, \ab( H\ab(\frac{\kappa+\sqrt{\lambda_0}z_0+\sqrt{\lambda_1}z_1}{\sqrt{\mu}}) )^m \,,
\end{equation}
where
\begin{align}
    \lambda_0 & = q_{0,\text{eff}}+\ev{g'}^2 \bar{d}_0 \,,                              \\
    \lambda_1 & = q_{1,\text{eff}}-q_{0,\text{eff}}+\ev{g'}^2 (\bar{d}_1-\bar{d}_0) \,, \\
    \mu       & = \sigma^2-q_{1,\text{eff}}+\ev{g'}^2 (\bar{c}-\bar{d}_1) \,.
\end{align}

Thus, the 1-RSB free entropy is given as follows:
\begin{align}
    \begin{split}
    f_\text{1-RSB} &= \extr_{\substack{m,q_1,q_0,\bar{c},\bar{d}_1,\bar{d}_0,\\\hat{E},\hat{q}_1,\hat{q}_0,\hat{c},\hat{d}_1,\hat{d}_0}} \Bigg\lbrace
    \frac{1}{2}\ab(\hat{E}+(1-m)q_1\hat{q}_1+mq_0\hat{q}_0+\bar{c}\hat{c}+(1-m)\bar{d}_1\hat{d}_1+m\bar{d}_0\hat{d}_0) \\
    & - \frac{1}{2} \Bigg( \frac{1}{m} \log \ab( \hat{E}+(1-m)\hat{q}_1+m\hat{q}_0-\hat{c}-(1-m)\hat{d}_1-m\hat{d}_0 ) \\
    & \qquad + \ab(1-\frac{1}{m}) \log \ab( \hat{E}+\hat{q}_1-\hat{c}-\hat{d}_1 ) - \frac{\hat{q}_0-\hat{d}_0}{ \hat{E}+(1-m)\hat{q}_1+m\hat{q}_0-\hat{c}-(1-m)\hat{d}_1-m\hat{d}_0 }  \Bigg) \\
    & + \frac{\alpha}{m} \int Dz_0 \, \log \int Dz_1 \, \ab( H\ab(\frac{\kappa+\sqrt{\lambda_0}z_0+\sqrt{\lambda_1}z_1}{\sqrt{\mu}}) )^m
    \Bigg\rbrace \,.
    \end{split}
    \label{eq:free_entropy_1rsb}
\end{align}

The saddle-point equations for $\bar{c},\bar{d}_1,\bar{d}_0$ give $\bar{c}=-1,\bar{d}_1=-q_1,\bar{d}_0=-q_0$.
Eliminating the order parameter $\hat{E},\hat{q}_1,\hat{q}_0,\hat{c},\hat{d}_1,\hat{d}_0$ gives
\begin{multline}
    f_\text{1-RSB} = \extr_{m,q_1,q_0} \Bigg\lbrace
    \frac{1}{2} \Bigg( \frac{1}{m} \log \ab( 1-(1-m)q_1-mq_0 ) + \ab(1-\frac{1}{m}) \log \ab( 1-q_1 ) \\
    + \frac{q_0}{1-(1-m)q_1-mq_0}  \Bigg) + \frac{\alpha}{m} \int Dz_0 \, \log \int Dz_1 \, \ab( H\ab(\frac{\kappa+\sqrt{\lambda_0}z_0+\sqrt{\lambda_1}z_1}{\sqrt{\mu}}) )^m
    \Bigg\rbrace \,,
    \label{eq:free_entropy_1rsb_simpified}
\end{multline}
where
\begin{align}
    \lambda_0 & = q_{0,\text{eff}}-\ev{g'}^2 q_0 \,,                        \\
    \lambda_1 & = q_{1,\text{eff}}-q_{0,\text{eff}}-\ev{g'}^2 (q_1-q_0) \,, \\
    \mu       & = \sigma^2-q_{1,\text{eff}}-\ev{g'}^2 (1-q_1) \,.
\end{align}

Next, we calculate the 1-RSB storage capacity $\alpha_\text{1-RSB}$.
As $\alpha \to \alpha_\text{1-RSB}$, we assume that $q_1\to 1,m \to 0$ with $r \coloneqq m/(1-q_1)$ approaching a finite value.
Let $q_1=1-\eps$ and we consider the limit $\eps\to 0$.
Here, we will follow the method of \cite{engel1992storage,zavatone2021activation}.
For the 1-RSB free entropy
\begin{equation}
    f_\text{1-RSB} = \extr_{r,\eps,q_0} \frac{1}{\eps} \ab\{ \eps G_s + \alpha \eps G_e\}
\end{equation}
to take a finite value as $\eps \to 0$, we need
\begin{equation}
    \min_{r,q_0} \lim_{\eps \to 0} \ab\{ \eps G_s + \alpha_\text{1-RSB} \eps G_e \} = 0 \,.
\end{equation}
Thus, it follows that
\begin{equation}
    \alpha_\text{1-RSB} = \min_{r,q_0} \ab\{ -\frac{\lim_{\eps \to 0} \eps G_s}{\lim_{\eps \to 0} \eps G_e} \} \,.
\end{equation}
The numerator is
\begin{align}
    \lim_{\eps\to 0} \eps G_s & = \lim_{\eps\to 0} \frac{1}{2} \ab( \frac{q_0}{1+r(1-q_0-\eps)} + \frac{1}{r} \log(\eps (1+r(1-q_0-\eps)) ) + \frac{r\eps-1}{r} \log \eps ) \\
                                      & = \frac{1}{2} \ab( \frac{q_0}{1+r(1-q_0)} + \frac{1}{r} \log( 1+r(1-q_0) ) ) \,.
\end{align}
The denominator is, using $\lambda_1 \to \sigma^2-q_{0,\text{eff}}-\ev{g'}^2 (1-q_0), \mu \to (\ev{g'^2} - \ev{g'}^2)\eps \eqqcolon \gamma \eps$,
\begin{align}
    \begin{split}
        \lim_{\eps\to 0} \eps G_e & = \lim_{\eps\to 0} \frac{1}{r}  \int Dz_0 \log \Bigg( \int_{-\infty}^{-Q} Dz_1                                                                                                                                                               \\
        & \quad + \int_{-Q}^\infty Dz_1 \ab( \frac{1}{\sqrt{2\pi}} \frac{\sqrt{\mu}}{\kappa+\sqrt{\lambda_0}z_0+\sqrt{\lambda_1}z_1} \exp\ab(-\frac{1}{2} \ab( \frac{\kappa+\sqrt{\lambda_0}z_0+\sqrt{\lambda_1}z_1}{\sqrt{\mu}})^2 ) )^{r\eps} \Bigg)
    \end{split} \\
     & = \frac{1}{r} \int Dz_0 \log\ab( \int_{-\infty}^{-Q} Dz_1 + \int_{-Q}^\infty Dz_1 \exp\ab(- \frac{r(\kappa+\sqrt{\lambda_0}z_0+\sqrt{\lambda_1}z_1)^2}{2\gamma} ) )                                                                                                                      \\
     & = \frac{1}{r} \int Dz_0 \log\ab( H(Q) + R \exp\ab(- \frac{r(\kappa+\sqrt{\lambda_0}z_0)^2}{2\gamma+r\lambda_1} ) H(-QR) ) \,,
\end{align}
where
\begin{equation}
    Q \coloneqq \frac{\kappa + \sqrt{\lambda_0} z_0}{\sqrt{\lambda_1}} \,,\quad R \coloneqq \sqrt{\frac{\gamma}{\gamma+r\lambda_1}} \,.
\end{equation}
Let
\begin{equation}
    \psi(r,q_0;\kappa) = -\int Dz_0 \log\ab( H(Q) + R \exp\ab(- \frac{r(\kappa+\sqrt{\lambda_0}z_0)^2}{2\gamma+r\lambda_1} ) H(-QR) ) \,.
\end{equation}
Then we obtain
\begin{equation}
    \alpha_\text{1-RSB} = \min_{r,q_0} \ab\{ \frac{1}{2\psi(r,q_0;\kappa)}\ab( \frac{rq_0}{1+r(1-q_0)} + \log( 1+r(1-q_0) ) ) \} \,.
    \label{eq:1rsb_cap}
\end{equation}

\subsection{Stability of the RS solution}

Let $\Delta q\coloneqq q_1-q_0,\,\Delta {\hat{q}}\coloneqq \hat{q}_1-\hat{q}_0,\,\Delta {\bar{d}}\coloneqq \bar{d}_1-\bar{d}_0,\,\Delta {\hat{d}}\coloneqq \hat{d}_1-\hat{d}_0$.
We seek the condition for $(\Delta q,\Delta {\hat{q}},\Delta {\bar{d}},\Delta {\hat{d}})=(0,0,0,0)$ to be a stable solution of the 1-RSB saddle-point equations.

The extremum condition of the 1-RSB free entropy \eqref{eq:free_entropy_1rsb} with respect to $\hat{q}_0,\hat{q}_1,\hat{d}_0,\hat{d}_1$ gives
\begin{align}
    q_0 & = \frac{\hat{q}_0 - \hat{d}_0}{(\hat{E}+(1-m)\hat{q}_1+m\hat{q}_0-\hat{c}-(1-m)\hat{d}_1-m\hat{d}_0)^2} \,, \\
    \begin{split}
        q_1 & = \frac{1}{m}\ab(\frac{1}{\hat{E}+(1-m)\hat{q}_1+m\hat{q}_0-\hat{c}-(1-m)\hat{d}_1-m\hat{d}_0} - \frac{1}{\hat{E}+\hat{q}_1-\hat{c}-\hat{d}_1} ) \\
        & \quad + \frac{\hat{q}_0-\hat{d}_0}{(\hat{E}+(1-m)\hat{q}_1+m\hat{q}_0-\hat{c}-(1-m)\hat{d}_1-m\hat{d}_0)^2} \,,
    \end{split}
\end{align}
\begin{equation}
    \bar{d}_0 = -q_0 \,,\quad \bar{d}_1 = -q_1 \,.
\end{equation}
When $\Delta {\hat{q}},\Delta {\hat{d}}\ll 1$,
\begin{align}
    \Delta q         & \approx \frac{1}{(\hat{E}+\hat{q}_1-\hat{c}-\hat{d}_1)^2} (\Delta {\hat{q}} - \Delta {\hat{d}}) \,,
    \label{eq:dat_delta_q}                                                                                             \\
    \Delta {\bar{d}} & = -\Delta q \,.
    \label{eq:dat_delta_d}
\end{align}

Next, we consider the extremum condition with respect to $q_0,q_1,d_0,d_1$.
Express the energy term as
\begin{align}
    G_e         & = \frac{1}{m}\int Dz_0 \log \int Dz_1 \exp(-m \mathcal{Y}) \,,                               \\
    \mathcal{Y} & = -\log H\ab(\frac{\kappa + \sqrt{\lambda_0}z_0+\sqrt{\lambda_1}z_1}{\sqrt{\mu}})             \\
                & = -\log \int Dx \, \Theta\ab(\kappa + \sqrt{\lambda_0}z_0+\sqrt{\lambda_1}z_1+\sqrt{\mu}x) \,.
\end{align}
Then we have
\begin{align}
    \diffp{\mathcal{Y}}{(\sqrt{\lambda_0} z_0)}  & = -\frac{1}{\int Dx \, \Theta} \int Dx \, \Theta' \,,                                                                                                                                                                       \\
    \frac{\partial^2 \mathcal{Y}}{\partial (\sqrt{\lambda_0} z_0)^2} & = -\frac{1}{\int Dx \, \Theta} \int Dx \, \Theta'' + \ab(\diffp{\mathcal{Y}}{(\sqrt{\lambda_0} z_0)})^2 \,,                                                                                                                 \\
    \diffp{\mathcal{Y}}{z_1}                     & = \sqrt{\lambda_1} \diffp{\mathcal{Y}}{(\sqrt{\lambda_0} z_0)} \,,                                                                                                                                                           \\
    \diffp{\mathcal{Y}}{q_0}                     & = q_{0,\text{eff}}'\ab(\frac{z_0}{2\sqrt{\lambda_0}} - \frac{z_1}{2\sqrt{\lambda_1}}) \diffp{\mathcal{Y}}{(\sqrt{\lambda_0} z_0)} \,,                                                                                        \\
    \diffp{\mathcal{Y}}{q_1}                     & = q_{1,\text{eff}}'\ab(\frac{z_1}{2\sqrt{\lambda_1}} \diffp{\mathcal{Y}}{(\sqrt{\lambda_0} z_0)} - \frac{1}{2} \ab(\frac{\partial^2 \mathcal{Y}}{\partial (\sqrt{\lambda_0} z_0)^2} - \ab(\diffp{\mathcal{Y}}{(\sqrt{\lambda_0} z_0)})^2)) \,.
\end{align}
Taking the derivative of $f_\text{1-RSB}$ with respect to $q_0$ gives
\begin{align}
    \diffp{f}{q_0} & = \frac{m}{2}\hat{q}_0 + \frac{\alpha}{m}\int Dz_0 \frac{1}{\int Dz_1 \exp(-m\mathcal{Y})} \int Dz_1 \exp(-m\mathcal{Y}) (-m) \diffp{\mathcal{Y}}{q_0} \\
    \begin{split}
        & = \frac{m}{2}\hat{q}_0-\alpha q_{0,\text{eff}}' \int Dz_0\frac{1}{\int Dz_1 \exp(-m\mathcal{Y})}                                                  \\
        & \qquad \int Dz_1 \exp(-m\mathcal{Y}) \ab(\frac{z_0}{2\sqrt{\lambda_0}} - \frac{z_1}{2\sqrt{\lambda_1}}) \diffp{\mathcal{Y}}{(\sqrt{\lambda_0} z_0)} \,.
    \end{split}
\end{align}
Integrating by parts with respect to $z_1$, we have
\begin{align}
    \begin{split}
        & \int Dz_1 \exp(-m\mathcal{Y}) \ab( -\frac{z_1}{2\sqrt{\lambda_1}}) \diffp{\mathcal{Y}}{(\sqrt{\lambda_0} z_0)}            \\
        & = -\frac{1}{2\sqrt{\lambda_1}} \int Dz_1 \diffp{}{z_1} \ab(\exp(-m\mathcal{Y})\diffp{\mathcal{Y}}{(\sqrt{\lambda_0}z_0)})
    \end{split} \\
     & = -\frac{1}{2\sqrt{\lambda_1}} \sqrt{\lambda_1}\diffp{}{{(\sqrt{\lambda_0}z_0)}} \int Dz_1 \exp(-m\mathcal{Y})\diffp{\mathcal{Y}}{(\sqrt{\lambda_0}z_0)}                 \\
     & = -\frac{1}{2} \diffp{}{(\sqrt{\lambda_0}z_0)} \int Dz_1 \exp(-m\mathcal{Y})\diffp{\mathcal{Y}}{(\sqrt{\lambda_0}z_0)} \,.
\end{align}
Integrating by parts with respect to $z_0$, we have
\begin{align}
    \begin{split}
        & \int Dz_0 \frac{1}{\int Dz_1 \exp(-m\mathcal{Y})} \int Dz_1 \exp(-m\mathcal{Y}) \frac{z_0}{2\sqrt{\lambda_0}} \diffp{\mathcal{Y}}{(\sqrt{\lambda_0}z_0)}                      \\
        & = \frac{1}{2} \int Dz_0 \diffp{}{(\sqrt{\lambda_0}z_0)} \ab(\frac{1}{\int Dz_1 \exp(-m\mathcal{Y})} \int Dz_1\exp(-m\mathcal{Y}) \diffp{\mathcal{Y}}{(\sqrt{\lambda_0}z_0)})
    \end{split} \\
    \begin{split}
        & = \frac{1}{2} \int Dz_0 \frac{1}{\int Dz_1 \exp(-m\mathcal{Y})} \diffp{}{(\sqrt{\lambda_0}z_0)} \int Dz_1\exp(-m\mathcal{Y}) \diffp{\mathcal{Y}}{(\sqrt{\lambda_0}z_0)} \\
        & \qquad + \frac{m}{2} \int Dz_0 \ab(\frac{1}{\int Dz_1 \exp(-m\mathcal{Y})} \int Dz_1\exp(-m\mathcal{Y}) \diffp{\mathcal{Y}}{(\sqrt{\lambda_0}z_0)})^2 \,.
    \end{split}
\end{align}
Thus we have
\begin{equation}
    \diffp{f}{q_0} = \frac{m}{2}\hat{q}_0  - \alpha \frac{m}{2} q_{0,\text{eff}}' \int Dz_0 \ab(\frac{1}{\int Dz_1 \exp(-m\mathcal{Y})} \int Dz_1\exp(-m\mathcal{Y}) \diffp{\mathcal{Y}}{(\sqrt{\lambda_0}z_0)})^2 \,.
\end{equation}
Next, taking the derivative of $f_\text{1-RSB}$ with respect to $q_1$ gives
\begin{align}
    \diffp{f}{q_1} & = \frac{1-m}{2}\hat{q}_1 + \frac{\alpha}{m}\int Dz_0 \frac{1}{\int Dz_1 \exp(-m\mathcal{Y})} \int Dz_1 \exp(-m\mathcal{Y}) (-m) \diffp{\mathcal{Y}}{q_1} \\
    \begin{split}
        & = \frac{1-m}{2}\hat{q}_1-\alpha q_{1,\text{eff}}' \int Dz_0\frac{1}{\int Dz_1 \exp(-m\mathcal{Y})} \int Dz_1 \exp(-m\mathcal{Y})                                                                        \\
        & \qquad \times \ab(\frac{z_1}{2\sqrt{\lambda_1}} \diffp{\mathcal{Y}}{(\sqrt{\lambda_0} z_0)} - \frac{1}{2} \ab(\frac{\partial^2 \mathcal{Y}}{\partial (\sqrt{\lambda_0} z_0)^2} - \ab(\diffp{\mathcal{Y}}{(\sqrt{\lambda_0} z_0)})^2)) \,.
    \end{split}
\end{align}
Integrating by parts with respect to $z_1$, we have
\begin{align}
    \begin{split}
        & \int Dz_1 \exp(-m\mathcal{Y}) \ab( \frac{z_1}{2\sqrt{\lambda_1}}) \diffp{\mathcal{Y}}{(\sqrt{\lambda_0} z_0)}          \\
        & = \frac{1}{2} \diffp{}{(\sqrt{\lambda_0}z_0)} \int Dz_1 \exp(-m\mathcal{Y})\diffp{\mathcal{Y}}{(\sqrt{\lambda_0}z_0)}
    \end{split} \\
     & = \frac{1}{2} \int Dz_1 \exp(-m\mathcal{Y})\ab( -m\ab(\diffp{\mathcal{Y}}{(\sqrt{\lambda_0}z_0)})^2 + \frac{\partial^2 \mathcal{Y}}{\partial (\sqrt{\lambda_0} z_0)^2}) \,.
\end{align}
Thus we obtain
\begin{equation}
    \diffp{f}{q_1} = \frac{1-m}{2}\hat{q}_1  - \alpha \frac{1-m}{2} \int Dz_0 \frac{1}{\int Dz_1 \exp(-m\mathcal{Y})} \int Dz_1\exp(-m\mathcal{Y}) \ab(\diffp{\mathcal{Y}}{(\sqrt{\lambda_0}z_0)})^2
\end{equation}
and
\begin{align}
    \begin{split}
        \Delta {\hat{q}} & = \alpha \int Dz_0 \, \Bigg( q_{1,\text{eff}}' \frac{1}{\int Dz_1 \exp(-m\mathcal{Y})} \int Dz_1\exp(-m\mathcal{Y}) \ab(\diffp{\mathcal{Y}}{(\sqrt{\lambda_0}z_0)})^2 \\
        & \qquad -q_{0,\text{eff}}' \ab(\frac{1}{\int Dz_1 \exp(-m\mathcal{Y})} \int Dz_1\exp(-m\mathcal{Y}) \diffp{\mathcal{Y}}{(\sqrt{\lambda_0}z_0)})^2 \Bigg)
    \end{split} \\
     & = \alpha \int Dz_0 \, \ab(q_{0,\text{eff}}' \var\ab[ \diffp{\mathcal{Y}}{(\sqrt{\lambda_0}z_0)} ] + \ab(q_{1,\text{eff}}' - q_{0,\text{eff}}')\E\ab[\ab(\diffp{\mathcal{Y}}{(\sqrt{\lambda_0}z_0)})^2]) \,.
\end{align}
Here, the expectation and variance are taken with respect to the measure $\frac{1}{\int Dz_1 \exp(-m\mathcal{Y})} Dz_1 \exp(-m\mathcal{Y})$.
Expanding $\diffp{\mathcal{Y}}{(\sqrt{\lambda_0}z_0)}$ around $\Delta q=0$, we get
\begin{gather}
    \diffp{\mathcal{Y}}{(\sqrt{\lambda_0}z_0)} \approx \diffp{\mathcal{Y_\text{RS}}}{(\sqrt{\lambda}z_0)} + \frac{\partial^2 \mathcal{Y}_\text{RS}}{\partial (\sqrt{\lambda} z_0)^2} \sqrt{ (q'_\text{eff} - \ev{g'}^2) \Delta q} \,z_1 \,, \\
    \mathcal{Y}_\text{RS} = -\log H\ab(\frac{\kappa + \sqrt{\lambda}z_0}{\sqrt{\mu}}) \,.
\end{gather}
Also, when $\Delta q \ll 1$, $\frac{1}{\int Dz_1 \exp(-m\mathcal{Y})} Dz_1 \exp(-m\mathcal{Y})\approx Dz_1$ holds and thus $\var[z_1] \approx 1$.
Therefore,
\begin{equation}
    \Delta {\hat{q}} \approx\alpha \int Dz_0 \, \ab(q_{\text{eff}}' (q'_\text{eff}-\ev{g'}^2) \ab( \frac{\partial^2 \mathcal{Y}_\text{RS}}{\partial (\sqrt{\lambda} z_0)^2})^2 + q_{\text{eff}}'' \ab(\diffp{\mathcal{Y_\text{RS}}}{(\sqrt{\lambda}z_0)})^2) \Delta q \,.
    \label{eq:dat_delta_qhat}
\end{equation}
Similarly,
\begin{equation}
    \Delta {\hat{d}} \approx\alpha \int Dz_0 \, \ev{g'}^2 (q'_\text{eff}-\ev{g'}^2) \ab( \frac{\partial^2 \mathcal{Y}_\text{RS}}{\partial (\sqrt{\lambda} z_0)^2})^2 \Delta q \,.
    \label{eq:dat_delta_dhat}
\end{equation}

The condition that $(0,0,0,0)$ is stable in the system of equations \eqref{eq:dat_delta_q},\eqref{eq:dat_delta_d},\eqref{eq:dat_delta_qhat}, and \eqref{eq:dat_delta_dhat} is
\begin{equation}
    \frac{1}{(\hat{E}+\hat{q}-\hat{c}-\hat{d})^2} \cdot \alpha \int Dz_0 \, \ab((q_{\text{eff}}' - \ev{g'}^2)^2\ab( \frac{\partial^2 \mathcal{Y}_\text{RS}}{\partial (\sqrt{\lambda} z_0)^2})^2 + q_{\text{eff}}'' \ab(\diffp{\mathcal{Y_\text{RS}}}{(\sqrt{\lambda}z_0)})^2) < 1 \,.
\end{equation}
Reorganizing, we obtain Eq. \eqref{eq:dat_instability}.

\end{document}